\documentclass[10pt,a4paper,twocolumn,superscriptaddress,aps,prd,longbibliography,nofootinbib]{revtex4-2}
\pdfoutput=1
\usepackage[latin1]{inputenc}

\usepackage{amsmath}
\usepackage{amsfonts}
\usepackage{amssymb}
\usepackage{bbold}
\usepackage{graphicx}
\usepackage[usenames,dvipsnames]{color}
\usepackage{float}
\usepackage{multirow}
\usepackage{soul}

\usepackage[nohyperlinks, printonlyused, nolist]{acronym}

\usepackage[range-units=single,separate-uncertainty]{siunitx}
\usepackage{verbatim}

\usepackage{blindtext}

\begin{document}

\author{Maximilian Bauernfeind}
\thanks{These authors have contributed equally}
\affiliation{Physikalisches Institut, Universit\"at W\"urzburg, D-97074 W\"urzburg, Germany}
\affiliation{W\"urzburg-Dresden Cluster of Excellence ct.qmat, Universit\"at W\"urzburg, D-97074 W\"urzburg, Germany}

\author{Jonas Erhardt}
\thanks{These authors have contributed equally}
\affiliation{Physikalisches Institut, Universit\"at W\"urzburg, D-97074 W\"urzburg, Germany}
\affiliation{W\"urzburg-Dresden Cluster of Excellence ct.qmat, Universit\"at W\"urzburg, D-97074 W\"urzburg, Germany}

\author{Philipp Eck}
\thanks{These authors have contributed equally}
\affiliation{W\"urzburg-Dresden Cluster of Excellence ct.qmat, Universit\"at W\"urzburg, D-97074 W\"urzburg, Germany}
\affiliation{Institut f\"ur Theoretische Physik und Astrophysik, Universit\"at W\"urzburg, D-97074 W\"urzburg, Germany}

\author{Pardeep K. Thakur}
	\affiliation{Diamond Light Source, Harwell Science and Innovation Campus, 
Didcot, OX11 0DE, United Kingdom}

\author{Judith Gabel}
	\affiliation{Diamond Light Source, Harwell Science and Innovation Campus, 
Didcot, OX11 0DE, United Kingdom}

\author{Tien-Lin Lee}
	\affiliation{Diamond Light Source, Harwell Science and Innovation Campus, 
Didcot, OX11 0DE, United Kingdom}

\author{J\"org Sch\"afer}
\affiliation{Physikalisches Institut, Universit\"at W\"urzburg, D-97074 W\"urzburg, Germany}
\affiliation{W\"urzburg-Dresden Cluster of Excellence ct.qmat, Universit\"at W\"urzburg, D-97074 W\"urzburg, Germany}

\author{Simon Moser}
\affiliation{Physikalisches Institut, Universit\"at W\"urzburg, D-97074 W\"urzburg, Germany}
\affiliation{W\"urzburg-Dresden Cluster of Excellence ct.qmat, Universit\"at W\"urzburg, D-97074 W\"urzburg, Germany}

\author{Domenico Di Sante}
\affiliation{Institut f\"ur Theoretische Physik und Astrophysik, Universit\"at W\"urzburg, D-97074 W\"urzburg, Germany}
\affiliation{Department of Physics and Astronomy, University of Bologna, 40127 Bologna, Italy}
\affiliation{Center for Computational Quantum Physics, Flatiron Institute, New York, 10010 NY, USA}

\author{Ralph Claessen}
\email{e-mail: claessen@physik.uni-wuerzburg.de}
\affiliation{Physikalisches Institut, Universit\"at W\"urzburg, D-97074 W\"urzburg, Germany}
\affiliation{W\"urzburg-Dresden Cluster of Excellence ct.qmat, Universit\"at W\"urzburg, D-97074 W\"urzburg, Germany}

\author{Giorgio Sangiovanni}
\email{e-mail: sangiovanni@physik.uni-wuerzburg.de}
\affiliation{W\"urzburg-Dresden Cluster of Excellence ct.qmat, Universit\"at W\"urzburg, D-97074 W\"urzburg, Germany}
\affiliation{Institut f\"ur Theoretische Physik und Astrophysik, Universit\"at W\"urzburg, D-97074 W\"urzburg, Germany}

\date{\today}


\title{Design and realization of topological Dirac fermions on a triangular lattice}
\maketitle

\begin{acronym}
\acro{2D}[2D]{two-dimensional}
\acro{DFT}[DFT]{density functional theory}
\acro{DOS}[DOS]{density of states}
\acro{VASP}[VASP]{Vienna ab initio simulation package}
\acro{PAW}[PAW]{projector-augmented-plane-wave}
\acro{ISB}[ISB]{inversion symmetry breaking}
\acro{SOC}[SOC]{spin-orbit coupling}
\acro{QSHI}[QSHI]{quantum spin Hall insulator}
\acro{TI}[TI]{trivial insulator}
\acro{OAM}[OAM]{orbital angular momentum}
\acro{BZ}[BZ]{Brillouin zone}

\acro{STM}[STM]{scanning tunneling microscopy}
\acro{STS}[STS]{scanning tunneling spectroscopy}
\acro{ZBA}[ZBA]{zero bias anomaly}
\acro{ARPES}[ARPES]{angle-resolved photoelectron spectroscopy}
\acro{XPS}[XPS]{X-ray photoelectron spectroscopy}
\acro{LEED}[LEED]{low-energy electron diffraction}
\acro{XSW}[XSW]{X-ray standing wave}
\acro{UHV}[UHV]{ultra-high vacuum}

\acro{EDC}[EDC]{energy distribution curve}
\acro{CEC}[CEC]{constant energy contour}

\acro{CCM}[CCM]{constant current mode}
\acro{CHM}[CHM]{constant height mode}
\acro{LDOS}[LDOS]{local density of states}
\acro{LT}[LT]{low-temperature}

\end{acronym}

\bigskip
\noindent
{\bf 
Large-gap quantum spin Hall insulators are promising materials for room-temperature applications based on Dirac fermions. Key to engineer the topologically non-trivial band ordering and sizable band gaps is strong spin-orbit interaction. Following Kane and Mele's original suggestion, one approach is to synthesize monolayers of heavy atoms with honeycomb coordination accommodated on templates with hexagonal symmetry. Yet, in the majority of cases, this recipe leads to triangular lattices, typically hosting metals or trivial insulators. Here, we conceive and realize ``indenene'', a triangular monolayer of indium on SiC exhibiting non-trivial valley physics driven by local spin-orbit coupling, which prevails over inversion-symmetry breaking terms. By means of tunneling microscopy of the 2D bulk we identify the quantum spin Hall phase of this triangular lattice and unveil how a hidden honeycomb connectivity emerges from interference patterns in Bloch $\boldsymbol{p_x\pm i p_y}$-derived wave functions.}


\begin{figure*}
    \centering
    \includegraphics[width=1\textwidth]{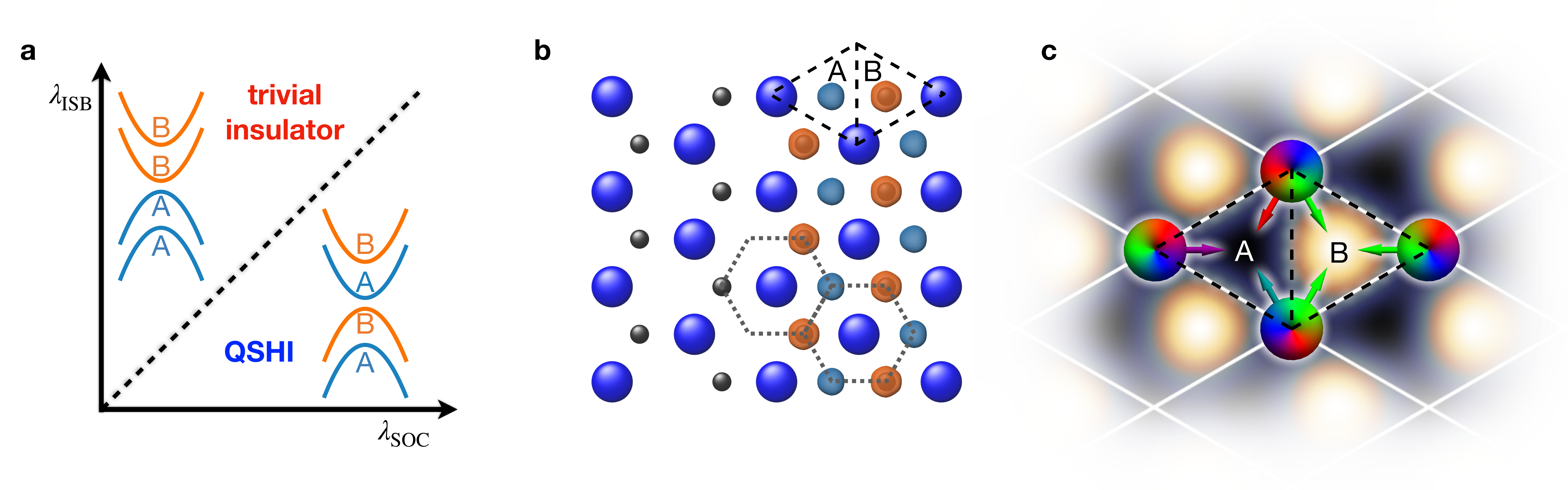}
    \caption{\textbf{Dirac fermions on a triangular lattice} -- 
    \textbf{a}, Sketch of the phase diagram as a function of  \acs{SOC} ($\lambda_\mathrm{SOC}$) and \ac{ISB} ($\lambda_\mathrm{ISB}$). The insets illustrate the band ordering of the Dirac fermions at the K and K$^\prime$ points in the \acs{QSHI} and trivial insulator phase. The coloring and the A/B labels correspond to the charge localization in \textbf{b}.
    \textbf{b}, Triangular lattice (blue spheres) with substrate-induced \ac{ISB}, as indicated by the C atoms (black spheres) in the first bilayer of the SiC(0001) substrate (see Fig.~\ref{fig:uc_characterization}a,c). The right half shows a honeycomb lattice (dotted gray lines) emerging from the charge localization of the chiral Dirac states, which peaks in the voids of the triangular lattice.
    \textbf{c}, Interference mechanism between Bloch and orbital phases determining the charge localization, shown schematically for the example of a $p_-$ orbital at the K point: the total phases (indicated by the rainbow color scheme) of neighboring sites contributing to the Bloch wave function interfere constructively at the B site. This promotes a high (bright) charge density centered around B, while destructive interference  suppresses the charge localization in the A triangle.
    }
    \label{fig:intro}
\end{figure*}

\bigskip
\noindent

The electronic wave functions of quantum spin Hall materials wind in momentum
space in a topologically distinct way from ordinary insulators, as described by
the corresponding $\mathbb{Z}_2$-invariant. The quantized transport via
spin-polarized boundary modes is protected by time-reversal symmetry, making
\acp{QSHI} technologically attractive
\cite{RevModPhys.82.3045,RevModPhys.83.1057}.
Ideal platforms are \ac{2D} honeycomb systems, as these naturally host massive Dirac fermions at the K/K$^\prime$ points in momentum space. As drawn in Fig.~\ref{fig:intro}a, \ac{SOC} opens a non-trivial gap, whereas inversion symmetry breaking (\acsu{ISB}) counteracts \ac{SOC} and favors the trivial phase
\cite{kane2005z}. To achieve room-temperature operability, a large gap is
essential. Graphene is, for example, a poor \ac{QSHI} because its \ac{SOC} arises from weak 2$^{\text{nd}}$-nearest neighbor hopping processes between out-of-plane $p_z$ orbitals \cite{kane2005quantum}. To improve on the gap size, \ac{2D}
materials made of heavier elements relying on local \ac{SOC} are hence superior
\cite{liu2011quantum,xu2013large,li2018theoretical,chen2018large,di2019towards}.
So far, this type of materials design has been achieved only in bismuthene, a
honeycomb system featuring purely planar bonding of its 6$p$ orbitals
\cite{reis2017bismuthene}.

The strategy of replacing carbon with heavier atoms faces two serious challenges. First,
in the unavoidable presence of a substrate high-$Z$ elements tend to form
buckled structures that are hostile to topology \cite{rivero2014stability}. Second, many
elements order preferentially in triangular rather than in honeycomb lattices
when deposited on hexagonal templates (see Supplementary Information I), with no experimental realization of a \ac{QSHI} phase hitherto \cite{WangQSHI2DTrigonal}. 
Synthesizing triangular \ac{QSHI} would therefore potentially accelerate the steps towards the first single-layer \ac{QSHI} device concept as, for instance, the growth process would profit from the simplicity of a non-bipartite lattice.

Here, we realize a triangular lattice of indium on SiC(0001) with topological
band inversion at the valley momenta K/K$^\prime$. 
In ``indenene'', \ac{SOC} arises locally from In 5$p$ orbitals and opens a gap between valence and conduction bands of about 100\,meV. 
A global gap is guaranteed by the presence of the substrate which induces an anticrossing of the bands derived from the indium $p_z$ orbital and the two planar $p_\pm \propto p_x \pm i p_y$ chiral orbitals, respectively \cite{petersen2000simple}. 
The valence and conduction bands are further spin-split at the valleys, as a consequence of the in-plane \ac{ISB} of SiC(0001). 
The C atom of the surface SiC layer renders the two halves of the indenene unit cell (labeled as ``A'' and ``B'' in Fig.~\ref{fig:intro}b and hereafter) inequivalent.

The splitting of the Dirac bands at K/K$^\prime$ allows to determine
the topological nature via a direct energy-resolved analysis of the bulk bands:
As illustrated in Fig.~\ref{fig:intro}c, the phases of the $p_\pm$-derived Bloch wave functions give rise to constructive and destructive
interference in A and B. The resulting charge localization at the voids of the
triangular lattice induces an emergent honeycomb connectivity (see dark blue and orange spots in Fig.~\ref{fig:intro}b). 
The energy ordering of these valley-states and the A/B character of the corresponding bulk wave function distinguishes the \ac{QSHI} from the trivial phase in an unambiguous way, as sketched in Fig.~\ref{fig:intro}a.
This therefore stands out as an example of a topological classification through the spatial symmetry of the electronic \emph{bulk} wave functions probed by \ac{STS}.


\section*{Theoretical model}


\begin{figure*}[t]
\includegraphics[width=1.0\textwidth]{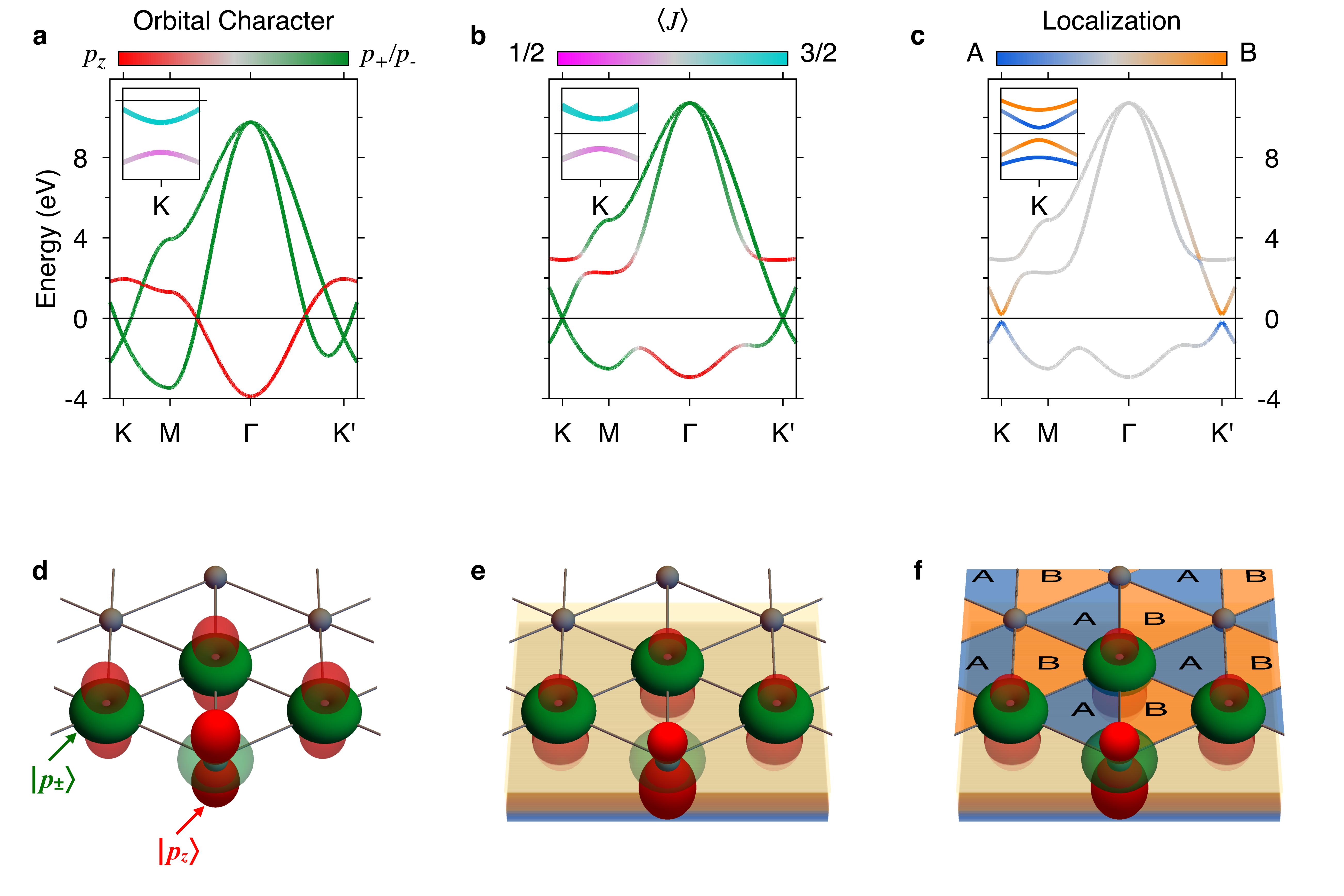}
\caption{\textbf{Model for a triangular QSHI on a substrate} -- \textbf{a-c}, Band structures
without (with) \acs{SOC} in the main panels (insets) and corresponding models \textbf{d-f} for a
$\{p_+,p_-,p_z\}$ basis on a triangular lattice. 
\textbf{a,d} The orthogonality of the $p_z$ and the in-plane orbitals promotes a metallic band structure with Dirac states at
K/K$^\prime$ which split in the presence of \acs{SOC} into states with strong
$J=\frac{1}{2}$ and $\frac{3}{2}$ character in valence and conduction band, 
respectively. 
\textbf{b,e} The presence of a substrate allows for $p_z$ and $p_\pm$ orbital-mixing and opens a hybridization gap. With \acs{SOC}, a topologically non-trivial
insulating ground state is realized (inset to \textbf{b}).
\textbf{c,f} in case of broken inversion symmetry, the blue (A) and orange (B) triangles become inequivalent. Without \acs{SOC} the system is in the trivial
phase and the whole valence (conduction) band localizes in the A- (B-)triangle. The inset to \textbf{c} shows the SOC-dominated non-trivial band ordering with spin-split $p_\pm$ states at the valley momenta. The valence (conduction) bands localize now on both triangles of the A/B sublattice. This distinctive A/B localization motif is the same at K and K$^\prime$.}
	\label{fig:model}
\end{figure*}

For a microscopic understanding of the physics of a triangular model of $p$ orbitals on a substrate we use the three spherical harmonics
\begin{equation*}
\left\{p_\pm\!=\!\frac{1}{\sqrt{2}}(p_x \pm ip_y) \,\, \hspace{3pt}, p_z
\right\}. 
\end{equation*}
In this basis, we step-wise introduce the key interactions relevant to stabilize the \ac{QSHI} phase with low-energy Dirac states at K/K$^\prime$, as illustrated in Fig.~\ref{fig:model}. 
The following tight-binding Hamiltonian captures the low-energy electronic structure of any realistic implementation, as we will see later in the \ac{DFT} calculations for indenene on SiC(0001).
The latter is the actual material realization that we propose here and our modeling allows us to precisely determine the conditions under which its \ac{QSHI} phase is realized.

In the freestanding triangular layer (Fig.~\ref{fig:model}a,d) the $D_{6h}$ point
symmetry yields Dirac-crossings of the $p_\pm$ in-plane orbitals at K/K$^\prime$
and prohibits the hybridization with the $p_z$ subspace resulting in a metallic
phase. Local (atomic) \ac{SOC} ($H^{\rm SOC}=\lambda_{\rm SOC}\vec{L}\cdot
\vec{S}$) becomes relevant at band touchings and opens a gap at K/K$^\prime$
between the $J=3/2$ and $1/2$ states of size $ \lambda_{\rm SOC}$ within the
in-plane subspace (see inset to Fig.~\ref{fig:model}a).

In the presence of a homogeneous substrate (Fig.~\ref{fig:model}b,e), the mirror
symmetry along the surface normal direction is broken and a hybridization gap opens between
the in- and out-of-plane orbitals. Considering \ac{SOC}, a non-trivial
insulating ground state is realized and the spin-degeneracy of the states with
mixed $p_\pm$ and $p_z$ orbital composition is lifted. This Rashba-like
splitting does not involve the Kramers doublet of the low-energy states
at K/K$^\prime$ that are protected by the $C_{6v}$ symmetry (see inset to
Fig.~\ref{fig:model}b).

Introducing a honeycomb substrate, such as SiC(0001) (see
Fig.~\ref{fig:intro}b and \ref{fig:uc_characterization}a), the symmetry is further
reduced to $C_{3v}$. As a consequence, \acs{ISB} renders the A and B
halves of the unit cell inequivalent, as schematically illustrated in
Fig.~\ref{fig:model}f. The corresponding non-local term, of strength $\lambda_{\rm ISB}$, acts within the in-plane subspace and opens a gap of size $3\sqrt{3}\lambda_{\rm ISB}$ at K/K$^\prime$.
$H^{\rm ISB}$ is diagonal in the spherical harmonics basis and promotes \ac{OAM}
polarization along the surface normal, competing with the topologically
non-trivial local \ac{SOC} gap (see inset to Fig.~\ref{fig:model}c). The $p_\pm$ valley-Hamiltonian reads
\begin{equation}
	\begin{split}
	H({\rm K/K^\prime})&=H^{\rm SOC}+H^{\rm ISB}({\rm K/K^\prime}) \nonumber\\
	&=\lambda_{\rm SOC}L_z\otimes S_z \pm\frac{3\sqrt{3}}{2}\lambda_{\rm ISB}L_z\otimes\mathbb{1}.
	\end{split}
\end{equation}
Depending on the relative strength of $\lambda_{\rm SOC}$ and $\lambda_{\rm
ISB}$, the gap at K/K$^\prime$ is dominated by either of the two types of
interaction, which defines the topological phase as indicated in Fig.~\ref{fig:intro}a. All cases shown in the
insets of Fig.~\ref{fig:model}a-c correspond to $ \lambda_{\rm SOC} >
3\sqrt{3}\lambda_{\rm ISB}$, i.e., to the topologically non-trivial band ordering. 
More details on the model can be found in Supplementary Information II. 

As mentioned above, the \acs{ISB} potential, whose strength depends on the substrate
and the bonding distance $d$, distinguishes between A and B (Fig.~\ref{fig:intro}b). Consequently, the
charge will tend to localize on the energetically lower triangle.
The arrows in Fig.~\ref{fig:intro}c sketch the interference mechanism between the lattice Bloch and the \ac{OAM} phases determining the charge-density profile (see Supplementary
Information II.E). In the trivial \acs{ISB}-driven phase, this interplay leads to
both spin-valence (conduction) bands at K and K$^\prime$ localizing in the A-
(B-)triangle -- see band structure without \ac{SOC} in the main panel of
Fig.~\ref{fig:model}c. The situation changes if the
\ac{SOC}-splitting dominates: the charge associated to the valence band doublet
is localized alternatingly in the A and B voids and the same is true for the two
unoccupied eigenvalues, as illustrated by the corresponding colors in the inset
to Fig.~\ref{fig:model}c.
A crucial observation is that the charge localization pattern is identical at both valley momenta, since the OAM polarization and the Bloch phase are odd under inversion.

An interference pattern of similar nature has been theoretically discussed in twisted bilayer graphene, though with lattice phases originating from the moir\'{e} superstructure, i.e. extending over much longer interatomic distances than in this case \cite{koshino2018maximally}.
Here, the A/B character represents an emergent honeycomb lattice degree of freedom, intimately linked to the topology of
the triangular $p$ model. 
Its role resembles that of the sublattice index in the graphene Kane-Mele Hamiltonian and it can be associated to the topological gap inversion. 
Further, it induces the (chirality-dependent) real-space localization of the bulk wave functions that can be measured directly in \ac{STM}.


\section*{Indenene on SiC(0001)}


\begin{figure*}[t]
	\includegraphics[width=1\textwidth,angle=0,clip=true]{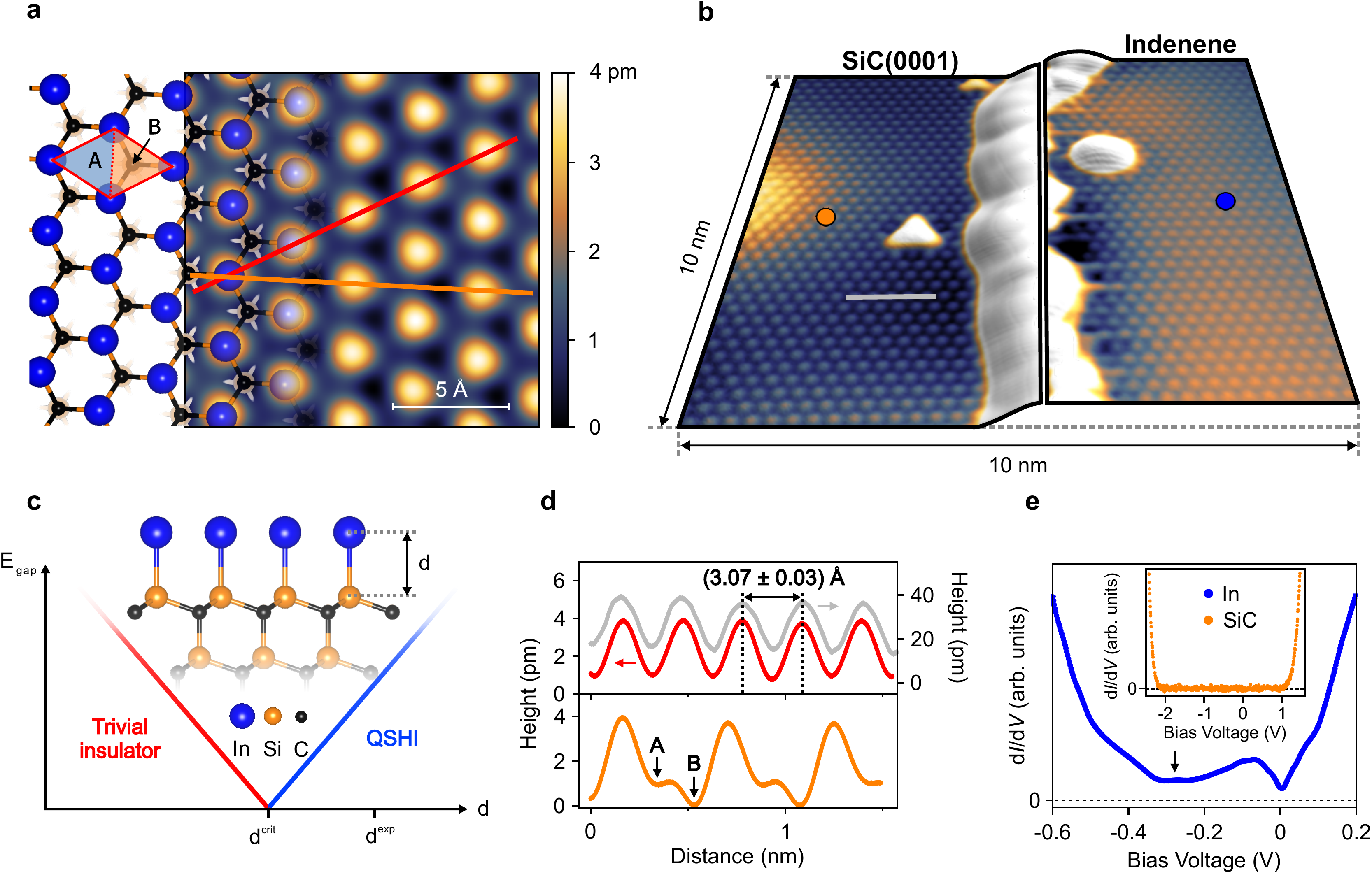}

\caption{
{\bf Triangular indium monolayer on SiC(0001)} -- 
\textbf{a},
\acs{STM} topography image (constant current mode with
$V_{\text{set}}=1.5\,\si{\volt}$ and
$I_{\text{set}}=50\,\si{\pico\ampere}$) of indenene next to the structural
model illustrating the triangular atom lattice forming on the SiC(0001) substrate.
Also shown is the $(1\times1)$ unit cell with its inequivalent A and B halves. 
At the chosen bias voltage \acs{STM} probes essentially the out-of-plane In $p_z$ orbitals
such that the bright spots directly reflect the positions of the In atoms.
\textbf{b},
Film edge between an indenene monolayer and uncovered SiC. Both lattices appear structurally identical but are
distinguishable by their electronic spectra in \textbf{e}. Imaging parameters
for SiC and indenene are ($V_{\text{set}}=1.65\,\si{\volt}$,
$I_{\text{set}}=80\,\si{\pico\ampere}$) and
($V_{\text{set}}=0.45\,\si{\volt}$,
$I_{\text{set}}=100\,\si{\pico\ampere}$), respectively.
\textbf{c}, 
Side view of the structural model highlighting the bond length $d$ between the
indenene layer and the Si-terminated 4H-SiC substrate. 
The accompanying graph schematically illustrates the evolution of the energy gap $E_{gap}$ as a function of $d$ in the vicinity of the topological phase transition at $d^{crit}$ (see text for further details).
\textbf{d}, 
\ac{STM} height profiles of indenene
along the red and orange 
paths in \textbf{a} 
showing the lattice constant and the asymmetry between the A and B voids of the unit cell, respectively. The gray height profile taken on the uncovered SiC(0001) substrate in \textbf{b} proves the identical lattice constants.
\textbf{e},
$\text{d}I/\text{d}V$ spectra measured on an indenene film and SiC substrate. In contrast to the metallic states in the close vicinity of the Fermi level in indenene, SiC exhibits a wide tunneling gap of approximately $3.2\,\si{eV}$ \cite{SeyllerBandGapSiC}. 
The black arrow indicates conductance minimum attributed to the Dirac point of indenene.
	}
	\label{fig:uc_characterization}
\end{figure*}


\begin{figure*}[!t]
    \vspace{5mm}
    \includegraphics[width=1\textwidth]{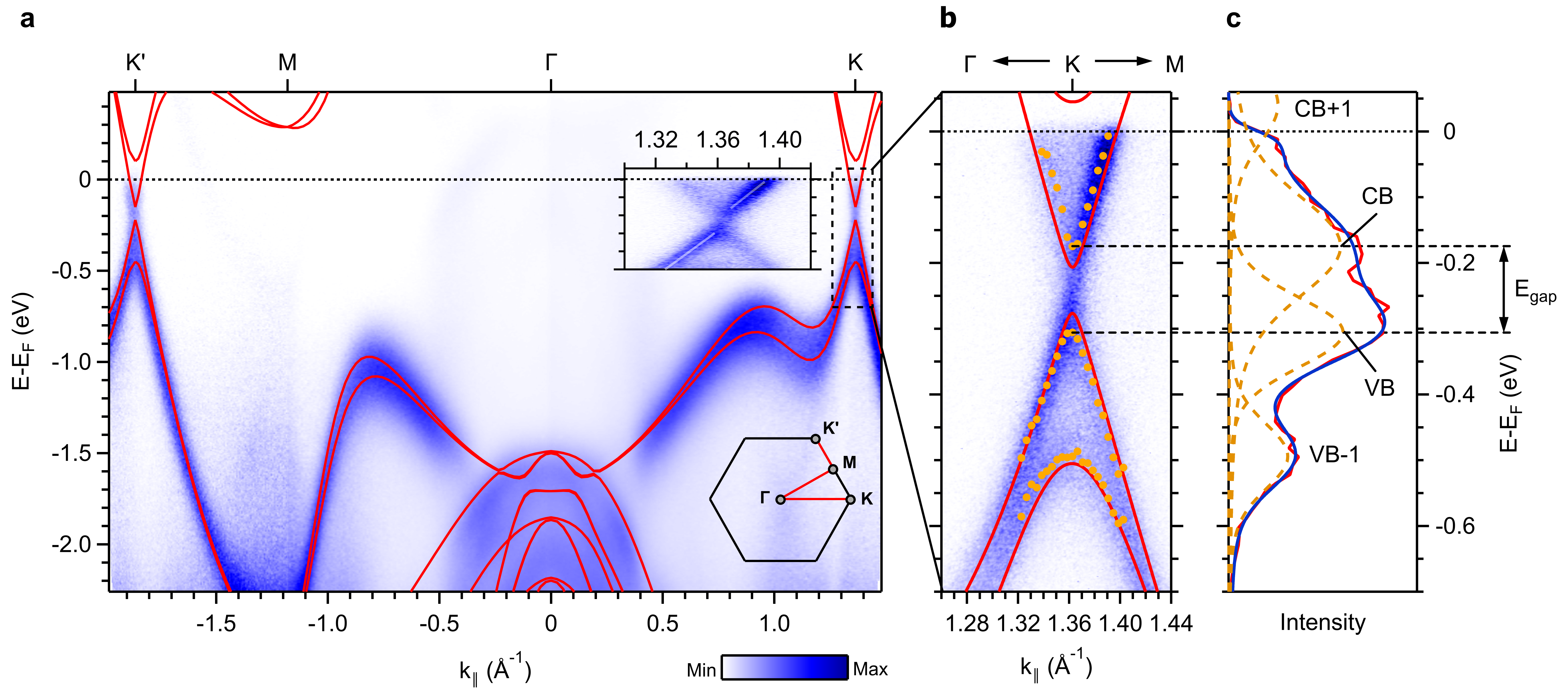}
\caption{{\bf Band structure of indenene} -- 
\textbf{a}, Comparison of \acs{ARPES} and \acs{DFT} band structure (red) and Brillouin zone schematics for the probed high symmetry lines. The
faint dispersive features around the Fermi level are artifacts originating from
satellite lines of the non-monochromatized He-I radiation (see Methods). The inset shows the
dispersive behavior at the K-point on a strongly enlarged momentum scale,
illustrating the deviation from an ungapped Dirac double-cone. \textbf{b},
Detailed analysis of the K-point dispersions. The orange markers indicate the
peak positions obtained from peak fitting of the \acsp{EDC} at selected
k-vectors. \textbf{c}, \acs{EDC} at the K point (red) and its peak fit (blue -
sum, orange dashed lines - peaks) accounting for the first two valence band
states (VB, VB-1) and the lowest conduction band (CB).
}
	\label{fig:ARPES_DFT_bands}
\end{figure*}

A monolayer of indium atoms deposited on a Si-terminated 4H-SiC(0001) is an ideal approach to attain a physical realization of our model.
Its synthesis is
achieved by molecular beam epitaxy, leading to high-quality indenene films as
characterized by standard surface science tools (see Methods and Supplementary Information III). Topographic imaging by STM confirms
the well-ordered triangular lattice formed by the In atoms as shown in
Fig.~\ref{fig:uc_characterization}a. According to the atomic arrangement obtained by total energy
minimization within \ac{DFT} (see methods) the In atoms bind directly to the uppermost silicon atoms of the
SiC substrate (T1 position), as depicted in Figs.~\ref{fig:uc_characterization}a
and c. This adsorption geometry translates into a $(1\times1)$ surface
periodicity of the indenene layer with respect to the SiC substrate, with
identical in-plane lattice constants as confirmed by the STM line profiles in
Fig.~\ref{fig:uc_characterization}d \cite{Stockmeier4HSiC}. Note in particular the asymmetric
height profile along the orange path (Fig.~\ref{fig:uc_characterization}d) which reflects the
\ac{ISB} imposed by the C atoms in the first bilayer of the SiC substrate. 

Figure~\ref{fig:uc_characterization}b shows an \ac{STM} image of an  
indenene layer next to uncovered SiC substrate.
Due to the identical triangular
lattice both surfaces appear structurally indistinguishable. Electronically, though,
both systems differ significantly in the differential
tunneling conductance $\text{d}I/\text{d}V$ (Fig.~\ref{fig:uc_characterization}e), a measure of the 
the \ac{LDOS}. While SiC displays the expected wide energy gap
\cite{SeyllerBandGapSiC}, we find for indenene finite spectral weight throughout
the entire probed energy region.

Before further analyzing the indenene \ac{LDOS}, we first turn to its momentum-resolved
electronic structure. The red curves in Fig.~\ref{fig:ARPES_DFT_bands}a represent the DFT
band structure for the fully relaxed indenene-substrate combination. Apart from
the substrate-related bands below $-1.5$\,eV at the center of the Brillouin zone,
all other bands are of In $p$ character and reproduce the features seen
in our tight-binding model. 
In particular, we observe a Dirac-like dispersion around K/K$^{\prime}$ with an additional spin splitting resulting in four distinct bands, as expected in the presence of \ac{SOC}. 
Correspondingly, a fundamental band gap of size $E_\text{gap}\! =\! 70$\,meV is present (note that the energy
scale in Fig.~\ref{fig:ARPES_DFT_bands} refers to the experimental Fermi level position;
\ac{DFT} \textit{per se} places $E_F$ in the gap).

Fig.~\ref{fig:ARPES_DFT_bands}a also shows the experimental band structure 
determined by \ac{ARPES}. It consists of well defined band features whose
dispersions are in remarkable agreement with the \ac{DFT} prediction. The only
notable deviation concerns the position of the Fermi level, which in the
experiment is shifted into the upper Dirac half-cone by $\approx250$\,meV due to
electronic charge transfer from the strongly $n-$doped substrate (see Methods
and Supplementary Information IV.A for details). This
extrinsic population of the conduction band minimum puts us into a position
to probe the band gap directly by ARPES. For this purpose
Fig.~\ref{fig:ARPES_DFT_bands}a shows a zoom-in of the gap region at the
K-point. Clearly the quasi-linear
dispersions of the upper and lower Dirac cones do not connect to each other (see
inset of Fig.~\ref{fig:ARPES_DFT_bands}a). A peak-fit of the energy distribution curves (EDCs) at the K-point in
Fig.~\ref{fig:ARPES_DFT_bands}c decomposes the spectrum into
three distinct peaks, namely the two first valence band states (denoted by VB-1
and VB) and the lowest conduction band state (CB). The next state (CB+1) is
essentially cut-off by the Fermi-Dirac function. Extending this decomposition to
selected $k$-vectors around the K-point yields the
orange markers in Fig.~\ref{fig:ARPES_DFT_bands}b and excellently traces the \ac{DFT} bands. Their smallest separation is indeed found at K,
yielding $E_\text{gap}\! \approx \!125\,\si{meV}$, in reasonable correspondence with the
\ac{DFT} value, considering that \ac{DFT} tends to underestimate band gaps.

With this information at hand, the minimum seen in the experimental \ac{LDOS}
around $-0.3\,$eV (arrow in Fig.~\ref{fig:uc_characterization}e) is readily identified as
the Dirac point, shifted to negative bias voltage due to the finite
$n$-doping. The fact that it appears only as a spectral dip rather than a truly
vanishing $\text{d}I/\text{d}V$ signal, is a characteristic consequence of the partially occupied CB \cite{FeenstrandopedSC,STSInSb(110),STSScN(001)} (see also Supplementary Information V.A).
We note in passing that our \ac{STS}
spectra show nm-scale spatial fluctuations of the chemical potential of the
order of $\pm40$\,meV (see Supplementary Information V.B),
presumably induced by inhomogeneities in doping concentration as known from
related semiconducting substrates \cite{WeidlichPRB,Zhang_2021}. In our \ac{ARPES} data this
effect will be spatially smeared out by the large photon spot (diameter $\approx
1$ mm) and contribute to the \ac{EDC} peak widths. Noteworthy, the indenene \ac{LDOS}
exhibits a second spectral depression that is always pinned at zero
bias, irrespective of local fluctuations. Various mechanisms have been
suggested as origin of such a \ac{ZBA} which, however, depend on the specific
probing details \cite{WeiteringCoulomb, PhononZBA, EfrosZBA}. It is therefore
not considered as an intrinsic feature of the electronic structure.

Having established the existence of a sizable band gap at the valley momenta,
we now address the question of its topological character. Our \ac{DFT}
calculation for the fully relaxed structure indicates a non-trivial phase
($\mathbb{Z}_2\!=\!1$) as derived from the \textit{ab initio} Wannier charge center
movement \cite{WCC} (see Supplementary Information II.G). Interestingly, as we have seen from our model, the
topology can be tuned by the relative strength of \ac{ISB} and \ac{SOC} (Fig.~\ref{fig:intro}a). 
For our particular case of
In/SiC the In-Si bond length $d$ turns out to be the relevant control parameter
(Fig.~\ref{fig:uc_characterization}c): The smaller the separation to the
substrate, the stronger will be the impact of substrate-induced ISB on the indenene layer. 
For small bond lengths this implies
$\lambda_\mathrm{ISB} \gg \lambda_\mathrm{SOC}$ (trivial band gap) whereas in
the opposite case the system is in the \ac{QSHI} phase. 
This picture is confirmed by
DFT for fixed (non-relaxed) bond lengths $d$, with the topological transition at
$d^{crit}\!=\!2.57\,\si{\angstrom}$. 
The equilibrium bonding distance for our In/SiC
system is $d^{DFT}\!=\!2.68\,\si{\angstrom}$ (see Methods) in excellent agreement
with the measured value of $d^{exp}\!=\!(2.67\pm0.04)\,\si{\angstrom}$
obtained by \ac{XSW} photoemission (see Supplementary
Information III.D).
From the distance we hence get a first, though indirect, hint that indenene is on the non-trivial  side  of  the  topological  phase  diagram. In the following, we present an unambiguous experimental determination of its topology, directly linked to the interference argument anticipated in Fig. 1c.


\section*{Topological classification}

\begin{figure*}[t]
	\includegraphics[width=1\textwidth]{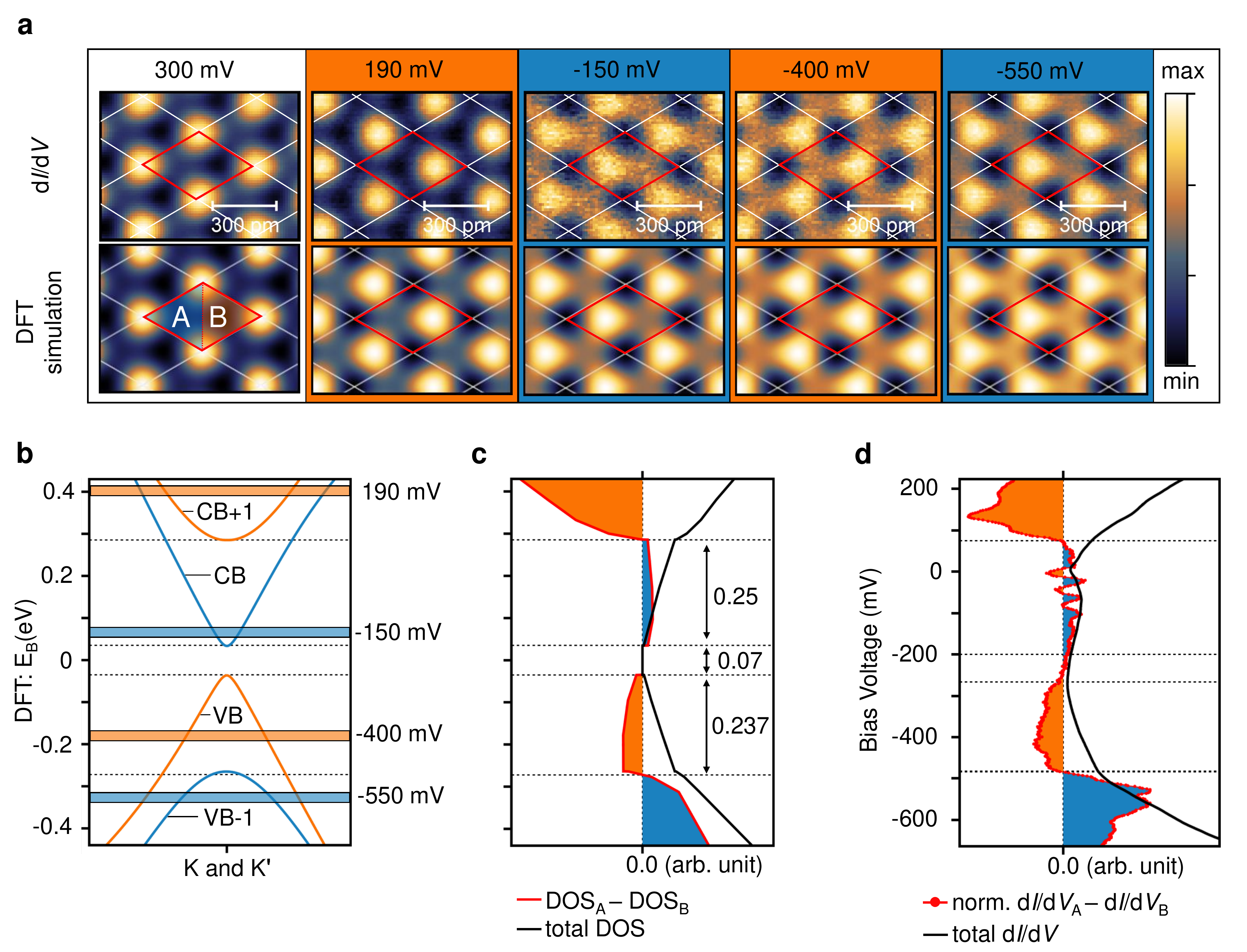}
	\caption{{\bf Non-trivial Dirac fermion charge localization} -- 
\textbf{a}, Spatially resolved constant height $\text{d}I/\text{d}V$ maps taken at
different energies in comparison with \acs{DFT} simulations. 
The In atom positions are calibrated by mapping the $p_z$ orbital dominated states at
$300\,\si{\milli\volt}$. Tuning the bias voltage to lower and eventually negative values 
results in distinct toggling of the \acs{LDOS} maximum between the two inequivalent A/B parts of the unit cell.
\textbf{b}, Close-up of the \acs{DFT} band structure and charge localization pattern at both valley momenta. The color code marks the
charge maximum at the A (blue) and B (orange)
sites. 
\textbf{c}, \acs{DOS} (black
line) and A/B difference (color line) from \ac{DFT}. Blue and orange areas correspond to
energies with charge maxima at site A and B, respectively. 
\textbf{d},
Differential conductance averaged over the entire unit cell and normalized
difference of the $\text{d}I/\text{d}V$ spectra
measured at the A/B sites. The noisy behavior near $E_F\!=\!0\,\si{\volt}$ is attributed
to the \acs{ZBA}. \Ac{STS} data are taken at a constant tip to sample distance
(see Methods for more details). Normalized difference spectra and $\text{d}I/\text{d}V$ maps
taken at various tip to sample distances agree qualitatively (see also Supplementary Information V.D, V.E).}
	\label{fig:topo_STM}
\end{figure*}

The chiral symmetry of the Dirac states on the triangular lattice can be exploited to access the topological nature directly from 2D bulk properties. 
Being composed of $p_\pm$ orbitals with defined \ac{OAM} the valley states assume an $e^{\pm i \phi}$ angular dependence around each atomic center. 
In combination with the Bloch phase picked up from one lattice site to the next the superposition of neighboring atomic orbitals causes characteristic interference effects, namely the
localization of the respective wavefunction at either the A or the B voids of
the unit cell, visualized in Fig.~\ref{fig:intro}c. As inferred from our model,
the actual information on the trivial \textit{vs}. inverted character of the
band gap is encrypted in the energy sequence of the A/B localization pattern of
the four valley states at K and K$^{\prime}$ (cf.~Fig.~\ref{fig:intro}a and inset of
Fig.~\ref{fig:model}c).

Experimentally, the energy-dependent charge distribution is best addressed by
\ac{STS}, probing the \ac{LDOS} with atomic resolution. The upper row of
Fig.~\ref{fig:topo_STM}a shows $\text{d}I/\text{d}V$ maps taken at selected bias voltages and
covering several unit cells. For comparison, the lower row shows the
corresponding \ac{DFT} simulations, accounting for the doping-induced Fermi
level shift between experiment and theory. Analogously to the topography map in
Fig.~\ref{fig:uc_characterization}a, we first calibrate the atomic positions and
lattice orientation by probing the indenene $p_z$-dominated states at
an experimental bias of $300\,\text{mV}$.

By lowering the tunneling voltage into the energy range of interest and with no
other contribution from elsewhere in the Brillouin zone, the \ac{STS} signal
becomes exclusively sensitive to the K/K$^{\prime}$ valley states. Indeed, at
$190\,\text{mV}$ the charge maximum has shifted away from the atomic center to
the B void of the unit cell. Tuning the bias to smaller and eventually negative
values ($-150\,\text{mV}$) leads to a switch of the charge localization,
now peaking at the A position. Going to even more negative bias voltage 
repeats the switching pattern, with the charge peak located at B and A 
at $-400$ and $-550\,\text{mV}$, respectively. 
Most importantly, the observed
alternation of charge localization is in excellent agreement with the corresponding
\ac{DFT} maps of the non-trivial system.

This behavior of the \ac{LDOS} can be further analyzed by directly comparing
the continuous energy dependence of the charge asymmetry between the A and B
halves of the unit cell (Figs.~\ref{fig:topo_STM}c and d for DFT and experiment,
respectively) to the valley band structure (Fig.~\ref{fig:topo_STM}b). Clearly,
 the charge difference switches sign each time a new valley state
contributes to the \ac{LDOS}. The noisy behavior of the experimental difference
spectrum around the Fermi level is attributed to the \ac{ZBA} in the total
$\text{d}I/\text{d}V$ curve (Fig.~\ref{fig:topo_STM}d) which tends to amplify small extrinsic
fluctuations in the A and B charge signals when taking their difference.
Overall, we find remarkable qualitative agreement between experiment and theory.
Specifically, an alternating ABAB charge localization sequence is established
when following the valley states in energy from VB-1 to CB+1, in clear
distinction from the AABB sequence predicted for a trivial insulator
(Fig.~\ref{fig:intro}a). 
Our \ac{STS} data thus confirm that indenene on SiC is a large-gap triangular \ac{QSHI}.

 
\section*{Outlook}
\label{Conclusion}

\noindent

From a general
perspective, the concept of an emerging honeycomb lattice in the interatomic voids of a
triangular atomic arrangement paves the way for the design of novel \ac{2D}
\acp{QSHI}. Our approach promotes in-plane chiral Dirac fermions at
K/K$^{\prime}$ whose mass term is determined by the interplay of local
\ac{SOC} and \ac{ISB}. These tunable electronic properties, combined with the
simple triangular geometry facilitating large-scale domain growth, are highly desirable
for room-temperature transport applications based on the utilization of topologically-protected edge states. 

The reported topological classification is achieved exclusively by means of local observables and it is intimately linked to the nature of the \emph{bulk} wave functions. For this reason, it represents an interesting complement to the common identification schemes based on the bulk-boundary correspondence. 
Its connection to the orbital angular momentum polarization in {\bf k}-space can be experimentally unveiled by exploiting the coupling of circularly polarized light to the orbital
magnetization and Berry curvature \cite{schuler2020local,unzelmann2020orbital}.
It also establishes fast early-stage material screening that is complementary to challenging
quantum transport experiments and can become relevant to topology beyond solid state physics,
e.g., in optical lattices of ultra-cold gases \cite{becker2010ultracold}.

\bigskip


\label{Methods}
{\noindent
	\textbf{Methods}
}

{\noindent
\small\textbf{Indenene synthesis, STM and photoemission measurements}\newline}
\footnotesize{4H-SiC(0001) samples ($12\,\si{mm}\times2.5\,\si{mm}$, \textit{n}-type doped (0.01 - 0.03) $\Omega$cm) with an atomically flat and well-ordered surface were prepared in a gaseous hydrogen dry-etching process \cite{SiCEtch, SeyllerHetchSiC}. Here, 2 slm H\textsubscript{2} and 2 slm He both with a purity of 7.0 were additionally filtered in gas purifiers and eventually introduced in a dedicated \ac{UHV} chamber with a pressure of approximately $950\,\si{mbar}$. The SiC sample was then etched at $1180^{\circ}\si{C}$ for 5 minutes. The smooth hydrogen passivated SiC sample \cite{SiCEtch, SeyllerHetchSiC} was then transferred \textit{in situ} to the epitaxy chamber where the surface quality was inspected with \ac{LEED} prior to the indium epitaxy. After a heating step which removed the H-saturation from the substrate, highly pure indium (99.9999$\%$) was evaporated from a standard Knudsen cell. Excessive indium was reduced thermally until only $(1\times1)$ \ac{LEED} diffraction spots remained (see Supplementary Information III.B).\\

\noindent{\ac{STM} data were acquired using a commercial Omicron \acl{LT} LT-\ac{STM} operated at $4.7\,\text{K}$ and a base pressure lower than $5\cdot10^{-11}\,\si{mbar}$. The chemically etched W-tip was conditioned and inspected on an Ag(111) crystal before and after measuring a sample. $\text{d}I/\text{d}V$ maps were taken at constant height using a standard lock-in technique with modulation frequency of $971\,\text{Hz}$ and modulation voltage of $V_{\text{rms}}=10\, \text{mV}$.
$\text{d}I/\text{d}V$ curves were recorded using the same lock-in technique. We achieved a semi-\ac{CHM} by interrupting the feedback loop at tunneling parameters with featureless topography in \ac{CCM} (e.g. at $I_{\text{set}}=50\,\si{\pico\ampere}$ and $V_{\text{set}}=-900\,\si{\milli\volt}$) followed by an approach of the tip to the sample surface by $\Delta z = -2.8\,\text{\AA}$ in order to generate a sufficiently large tunneling signal.
}\\

\noindent{\ac{ARPES} and \ac{XPS} data were recorded in our home-lab photoemission setup from Specs equipped with a hemispherical analyzer (PHOIBOS 100), a He-VUV lamp (UVS 300) generating photons of $21.2\,\text{eV}$, and a 6-axis LHe-cooled manipulator ($20\,\text{K}$ for \ac{ARPES}, room temperature for \ac{XPS}). The base pressure of this \ac{UHV} setup lies below $1\cdot10^{-10}\,\si{mbar}$. During LHe-cooled measurements the He partial pressure of the differential pumped He-VUV lamp did not exceed $1\cdot10^{-9}\,\text{mbar}$ in the \ac{UHV} chamber.} 
\\

\noindent{Room and low temperature \ac{XSW} measurements were performed at beamline I09 at Diamond Light Source in \ac{UHV} environment. The samples were prepared and characterized by \ac{ARPES} in our home-lab before shipping them \textit{in situ} in a \ac{UHV} suitcase with base pressure below $1\cdot10^{-9}\,\text{mbar}$. For more details, see Supplementary Information III.D.}}\\

{\noindent
\small\textbf{DFT calculations}\newline}
\footnotesize{For our theoretical study of indium on SiC(0001) we employed state-of-the-art first-principles calculations based on the density functional theory as implemented in the \ac{VASP} \cite{VASP1}, within the \ac{PAW} method \cite{VASP2,PAW}. For the exchange-correlation potential the HSE06 functional was used \cite{HSE06}, by expanding the Kohn-Sham wave functions into plane-waves up to an energy cut-off of 500 eV. We sampled the Brillouin zone on an $12\times12\times1$ regular mesh, and when considered, \ac{SOC} was self-consistently included \cite{SOC_VASP}. The energy decomposed densities are calculated on refined k-grids with a sampling of at least $90\times90\times1$ and $54\times54\times1$ for the low-energy states at K and at M, respectively, by selecting all relevant k-points with states inside the investigated energy window with the help of a Wannier Hamiltonian. The indenene low-energy models are extracted by projecting onto In $p$- and SiC $sp_3$-like functions (MLWF) by using the WANNIER90 package \cite{WANNIER90} to compute the $\mathbb{Z}_2$ topological invariant by following the general method of Soluyanov and Vanderbilt \cite{WCC}. We consider a $(1\times1)$ reconstruction of triangular In on four layers of Si-terminated SiC(0001) with an in-plane lattice constant of $3.07\,\text{\AA}$. The equilibrium structure is obtained by relaxing all atoms until all forces converged below $0.005\,\text{eV/\AA}$ resulting in an In-SiC distance of $d_{In-SiC}=2.68\,\text{\AA}$. To disentangle the electronic states of both surfaces a vacuum distance of at least 25 $\text{\AA}$ between periodic replicas in $z$-direction is assumed and the dangling bonds of the substrate terminated surface are saturated by hydrogen.}\\

{\noindent
\small\textbf{Tight-binding model}\newline}
\footnotesize{We consider a triangular lattice with a (In) $p$ basis
with a nearest neighbor interaction given by Slater-Koster Parameters \cite{slater1954simplified}. The on-site energies and transfer integrals 
are extracted from a Wannier Hamiltonian. Detailed information on the model can be found in Supplementary Information II.}

{\noindent
	\textbf{Data Availability}
The data that support the plots within this paper and other findings of this study are available from the corresponding author upon reasonable request.
}
\\


{\noindent
	\textbf{Acknowledgements}
We acknowledge Diamond Light Source for time on beamline I09 under proposals NT26419-1 and SI25151-4. The research leading to these results has received funding from the European Union's Horizon 2020 research and innovation programme under the Marie Sk\l{}odowska-Curie Grant Agreement No. 897276. We gratefully acknowledge the Gauss Centre for Supercomputing e.V. (www.gauss-centre.eu) for funding this project by providing computing time on the GCS Supercomputer SuperMUC-NG at Leibniz Supercomputing Centre (www.lrz.de). The authors are grateful for funding support from the
Deutsche Forschungsgemeinschaft (DFG, German Research
Foundation) under Germany's Excellence Strategy
through the W\"urzburg-Dresden Cluster of Excellence on
Complexity and Topology in Quantum Matter ct.qmat
(EXC 2147, Project ID 390858490) as well as through
the Collaborative Research Center SFB 1170 ToCoTronics
(Project ID 258499086).
}
\\

{\noindent
	\textbf{Author contributions}
M.B. and J.E. have realized the epitaxial growth and surface characterization and carried out the ARPES and STM experiments and their analysis. 
P.E. has conceived the theoretical ideas and performed the DFT, Wannier and Berryology calculations.
On the experimental side, contributions came from P.K.T., J.G., T.-L.L., J.S., S.M. and R.C., while D.D.S. and G.S. gave inputs to the theoretical aspects.
R.C. and G.S. supervised this joint project and wrote the manuscript together with all other authors.
}
\\

\begin{scriptsize}
\end{scriptsize}

\end{document}